\documentclass[10pt,conference]{IEEEtran}
\IEEEoverridecommandlockouts
\usepackage[T1]{fontenc}
\usepackage{cite}
\usepackage{amsmath,amssymb,amsfonts}
\usepackage{algorithmic}
\usepackage{graphicx}
\usepackage{textcomp}
\usepackage{xcolor}
\def\BibTeX{{\rm B\kern-.05em{\sc i\kern-.025em b}\kern-.08em
    T\kern-.1667em\lower.7ex\hbox{E}\kern-.125emX}}

\usepackage{balance}
\usepackage{bigstrut}
\usepackage{xcolor}
\usepackage{makecell} 
\usepackage{siunitx} 

\usepackage[hyphens]{url}
\usepackage[hidelinks]{hyperref}
\usepackage{graphicx}
\usepackage{tabularx}
\usepackage[most]{tcolorbox}
\usepackage{adjustbox}
\usepackage{listings}
\usepackage{enumitem}
\usepackage{amsmath}
\usepackage{bigstrut}
\usepackage{sepnum}
\usepackage{caption}
\usepackage[linesnumbered,ruled,noend]{algorithm2e}
\usepackage{subcaption}
\usepackage{multirow}
\usepackage{multicol}
\usepackage[framemethod=TikZ]{mdframed}
\usepackage{soul}
\usepackage{float}
\usepackage{pifont}
\usepackage{booktabs}
\usepackage{mathpartir}
\usepackage{makecell}

\usepackage{colortbl}
\definecolor{shaderow}{gray}{0.92}

\SetKw{Break}{break}
\SetKw{Continue}{continue}
\SetKw{Yield}{yield}
\newcounter{example}

\newcounter{boxedd}

\setlist[itemize]{noitemsep, leftmargin=1em}
\setlist[enumerate]{noitemsep, leftmargin=1em}
\newcommand{\nnum}[1]{\sepnum{.}{,}{}{#1}}

\setlist[itemize]{noitemsep, leftmargin=1em}
\setlist[enumerate]{noitemsep, leftmargin=1em}

\definecolor{dkred}{RGB}{87,10,10}
\definecolor{burntorange}{rgb}{0.8, 0.33, 0.0}
\definecolor{OrangeRed}{rgb}{1.0, 0.27, 0.0}
\definecolor{ForestGreen}{rgb}{0.0, 0.27, 0.13}

\definecolor{groovyblue}{HTML}{0000A0}
\definecolor{dkgreen}{HTML}{008000}
\definecolor{darkgray}{rgb}{.4,.4,.4}
\definecolor{bittersweet}{rgb}{1.0, 0.44, 0.37}

\newcommand{\cmark}{\textcolor{dkgreen}{\ding{51}}} 
\newcommand{\xmark}{\textcolor{red}{\ding{55}}}   

\SetCommentSty{mycommfont}

\usepackage{tikz}
\usetikzlibrary{arrows.meta, calc}

\lstdefinelanguage{JavaScript}{
  keywords={typeof, new, true, false, catch, function, return, null, catch, switch, var, if, in, while, do, else, case, break, const, let, require, import},
  keywordstyle=\color{blue}\bfseries,
  ndkeywords={class, export, boolean, throw, implements, import, this},
  ndkeywordstyle={\color[rgb]{0,0,0}\bfseries},
  keywordstyle={\color[rgb]{0,0,0}\bfseries},
  sensitive=true,
  stringstyle={\color[rgb]{0,0,1}\ttfamily},
  comment=[l]{//},
  morecomment=[s]{/*}{*/},
  commentstyle=\color[rgb]{0,0.5,0}\ttfamily,
  morestring=[b]',
  morestring=[b]"
}

\newcommand{\tool}{{\sc PyXSieve}}
\newcommand{\insight}{{\sc Insight}}
\newcommand{\lares}{{\sc Lares}}
\newcommand{\xray}{{\sc PyXray}}

\newcommand{\empirical}[1]{#1}

\newcommand{\point}[1]{{\noindent\bf #1:} }

\DeclareRobustCommand{\calloutfigure}[1]{%
  \raisebox{1pt}{%
    \tikz[baseline=(char.base)]{%
      \node[draw=black, fill=white, circle,
            inner sep=1.0pt, font=\scriptsize\bfseries]
            (char) {#1};}}}

\newcommand{\callout}[1]{%
  \raisebox{1pt}{%
    \tikz[baseline=(char.base)]{%
      \node[draw=black, fill=white, circle,
            inner sep=1.0pt, font=\scriptsize\bfseries]
            (char) {#1};}}}
\newcommand{\calloutt}[1]{%
  \raisebox{1pt}{%
    \tikz[baseline=(char.base)]{%
      \node[draw=black, fill=white, circle,
            inner sep=0.5pt, font=\scriptsize\bfseries]
            (char) {#1};}}}

\definecolor{rqaccent}{RGB}{31,78,121}
\definecolor{rqback}{RGB}{240,244,249}
 
\newtcolorbox{takeaway}[1]{
  enhanced, breakable,
  colback=rqback, colframe=rqaccent,
  boxrule=0.5pt, arc=2pt, outer arc=2pt,
  left=3pt, right=3pt, top=2pt, bottom=2pt,
  fonttitle=\bfseries\scriptsize, coltitle=white,
  colbacktitle=rqaccent,
  title={#1}, toptitle=0pt, bottomtitle=0pt,
  fontupper=\small,
  before skip=5pt, after skip=5pt,
}


\begin{document}
\sloppy

\title{Cross-Ecosystem Vulnerability Analysis for Python Applications}

\DeclareRobustCommand*{\IEEEauthorrefmark}[1]{%
  \raisebox{0pt}[0pt][0pt]{\textsuperscript{\footnotesize #1}}}

\author{%
\IEEEauthorblockN{%
Georgios Alexopoulos\IEEEauthorrefmark{1,2},
Nikolaos Alexopoulos\IEEEauthorrefmark{3},
Thodoris Sotiropoulos\IEEEauthorrefmark{4},\\
Charalambos Mitropoulos\IEEEauthorrefmark{5},
Zhendong Su\IEEEauthorrefmark{4},
Dimitris Mitropoulos\IEEEauthorrefmark{1,2}}
\IEEEauthorblockA{%
\IEEEauthorrefmark{1}University of Athens, Greece \quad
\IEEEauthorrefmark{2}National Infrastructures for Research and Technology, Greece\\
\IEEEauthorrefmark{3}Athens University of Economics and Business, Greece \quad
\IEEEauthorrefmark{4}ETH Zurich, Switzerland\\
\IEEEauthorrefmark{5}Technical University of Crete, Greece\\[2pt]
\{grgalex, dimitro\}@ba.uoa.gr \quad
alexopoulos@aueb.gr \quad
cmitropoulos@tuc.gr\\
\{theodoros.sotiropoulos, zhendong.su\}@inf.ethz.ch}%
}\maketitle

\begin{abstract}
Python packages often bundle
third-party native libraries (e.g.\ OpenSSL)
within their distributions,
a practice known as~\emph{vendoring}.
A vendored library is either copied from
an OS distribution package
or compiled from the upstream project's source.
Given a known vulnerability in such a library,
identifying the affected Python packages
is a challenging problem that spans both the Python
and OS ecosystems.
Current vulnerability scanners
either ignore vendored libraries entirely,
producing false negatives,
or ignore their origin,
producing false positives by missing the backported fixes
that OS distributions apply.

Our key insight is that accurate
cross-ecosystem vulnerability analysis requires
the exact~\emph{provenance} of each vendored binary:
not only its upstream project and version,
but the specific OS package and version,
or,
when built from source,
the exact upstream release.
We recover provenance deterministically rather
than by similarity-based estimation,
relying on two key observations.
First,
for binaries copied from OS packages,
Python packaging tools rewrite binary metadata
but leave its executable code unchanged.
We hash this unchanged code and match it against
a database of historical OS package artifacts.
Second,
libraries built from source typically expose
their version through callable interfaces.
We generate and apply per-library dynamic analysis rules
that read the version directly from the binary. 

On~\nnum{1878} Python packages with vendored native libraries,
we resolve exact provenance in~\emph{73.4\%} of cases overall,
and in \empirical{94.9\%} of the practically relevant cases
where at least one CVE exists for the library.
Our approach significantly outperforms
existing provenance-unaware approaches
on identifying whether vendored libraries are affected by known CVEs.
We integrate our approach with
existing Python and binary call graph generators
and perform the first
cross-ecosystem reachability analysis for vulnerabilities
in libraries vendored in Python packages.
Across the latest versions of the top~\nnum{100000} Python packages
and a sample of 10 CVEs in native libraries,
we identify 39 directly vulnerable packages
(47M+ monthly downloads) and 312 transitively affected packages.
We disclosed all findings to maintainers
and 54 have been fixed to date.

\end{abstract}


\section{Introduction}

\begin{figure*}[t]
  \begin{center}
    \includegraphics[width=\textwidth]{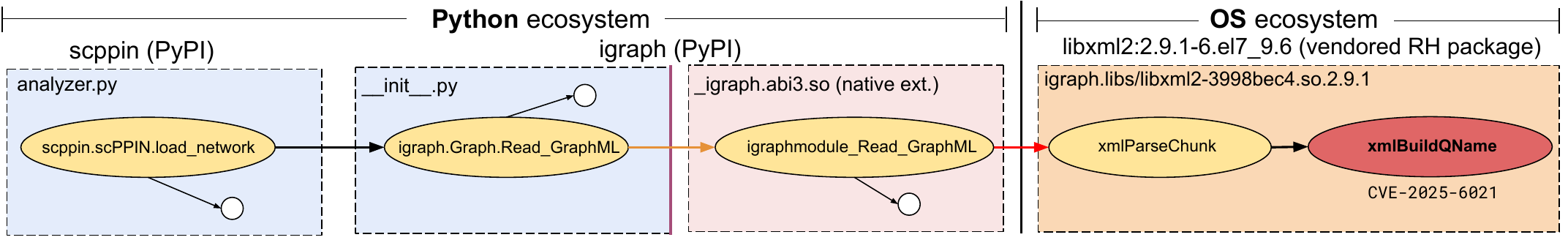}
  \end{center}
\caption{
Reachability of vulnerable binary functions from the \texttt{scppin} Python
package installed via \texttt{pip} on a Debian host.
}
\label{fig:run_ex}
\end{figure*}

Dependency management is critical
for software security and has
motivated extensive work on
Software Composition Analysis
(SCA)~\cite{JYTNWZ23,ZCXLZWSL23,JAHTNWZ24}.
Python,
one of the most popular programming
languages~\cite{tiobe-index},
presents unique challenges for
dependency analysis due to its hybrid nature.
Python applications depend not only on other
Python packages~\cite{bloat-study}
but also on native code.
This includes~\textit{native extensions}~\cite{pyxray,charon}
(Python modules implemented in C/C++/Rust
and distributed as {\tt .so} files)
as well as
third-party shared libraries~\cite{insight}.
Such libraries may be~\textit{vendored},
i.e.,
bundled within
a Python package distribution~\cite{pep513,auditwheel}
or provided by the host operating system (OS).
This creates a complex \emph{cross-ecosystem} dependency structure,
where numerous variants of the same shared library
may simultaneously exist on the system
and be invisible to
OS package managers~\cite{gorny2021,lwn-vendoring}.
Our motivating study on the top~\nnum{100000} PyPI packages shows
that vendoring is rare at source but pervasive in effect.
Although only~\nnum{1878} packages ($\sim$2\%) vendor shared libraries directly,
they affect~\empirical{28\%} of PyPI
through transitive dependencies.
When vulnerabilities are discovered in such libraries,
identifying the affected Python packages
becomes a cross-ecosystem analysis problem.

\point{Example}
Consider the scenario of Figure~\ref{fig:run_ex},
which we use as a running example
throughout the paper.
The Python package \texttt{scppin} is
installed on a Debian 13 host via \texttt{pip install}.
This package depends on the \texttt{igraph} Python package,
which ships a native extension
that calls {\tt libxml},
a shared library {\tt igraph} vendors by
bundling it into its distribution.
This forms a cross-ecosystem dependency chain.
This is because the {\tt libxml2} copy bundled in igraph originates
from a RHEL/CentOS package
and is vulnerable to~\href{https://nvd.nist.gov/vuln/detail/cve-2025-6021}{CVE-2025-6021}.

\point{The problem}
Our work aims to answer the following question:
\textit{is the scppin package affected by this CVE?}
A natural approach is function-level
reachability analysis~\cite{pyxray,bloat-study}.
As Figure~\ref{fig:run_ex} shows,
there is a complete call chain from {\tt scppin}
through {\tt igraph} into the function of interest {\tt xmlBuildQName}.
However,
reachability alone does not mean that {\tt scppin} is affected.
This is because {\tt xmlBuildQName} is reached in every copy of {\tt libxml2},
either a vulnerable or a patched one.
Therefore, exposure depends on whether the specific copy bundled inside
{\tt igraph} carries the fix, i.e., on the~\emph{binary's provenance}:
the exact version and patch level of the {\tt libxml2} copy.

\point{The gap}
Several techniques partially address
this binary provenance problem,
but none fully resolves it.
Similarity-based version
identification~\cite{LibvDiff,insight,JAHTNWZ24,b2sfinder}
and pattern-matching on embedded version strings~\cite{payer2021librarian}
recover at best an upstream version.
However,
neither identifies provenance,
i.e.,
the OS package a binary came from,
and therefore,
both miss backported patches entirely.
Patch-presence testing~\cite{Lares} is orthogonal:
it verifies whether a particular fix is present,
but only once you already know which vulnerability
and fixing commit to test against.
The fixing commit is often unavailable or incomplete
and identifying which vulnerabilities to check
is part of what provenance provides.
To the best of our knowledge,
no reliable,
automated approach exists that,
given an instance of a shared library (.so),
\textit{recovers its exact provenance,
i.e., the precise OS package and version,
or upstream release,
it originates from,
and thereby decides which known vulnerabilities actually affect it}.

%
%
%
%

\point{Approach}
To resolve provenance,
we rely on two key observations
of the Python ecosystem.
First,
Python's packaging pipeline for vendored libraries
rewrites only a binary's ELF metadata
(e.g., {\tt RPATH} and {\tt DT\_NEEDED} entries),
while leaving its code and data section unmodified.
Based on this,
our approach computes the hash of the untouched contents
and matches this hash against a
database of historical OS package binaries.
A match identifies the exact OS package
and version the vendored library came from.
When no match is found,
our approach treats the binary as built from upstream source,
a conjecture we empirically validate
(Section~\ref{sec:evaluation}).
Second,
most libraries expose their version at runtime
through callable APIs
(e.g., {\tt FT\_Library\_Version()}).
For binaries built from upstream source,
we exploit this complementary observation
and characterize all version-reporting mechanisms
through a taxonomy
across 50 frequently
vendored libraries (Table~\ref{tab:rules}).
We then use it to guide
lightweight version resolution rules.

\point{Results}
Evaluated on~\nnum{1878} Python packages
with vendored libraries,
our approach successfully resolves
the provenance in~\empirical{94.9\%} of security-relevant cases
(at least one CVE reported for the library),
identifying either the exact OS distribution package
and version
(including backported fixes) or,
when the library is not sourced from an OS package,
the upstream project and release version
from which the binary was built.

\point{Impact}
Our evaluation further showcases how
our provenance identification approach
enables cross-ecosystem vulnerability analysis.
Specifically,
we integrate it with state-of-the-art call graph
generators~\cite{pycg,ghidra,pyxray}
into a practical pipeline for assessing
the impact of vulnerabilities in third-party shared libraries
of Python packages and applications.
Our empirical study on the propagation of 10 CVEs
affecting popular native libraries
across the top-\nnum{100000} Python packages
identifies:
(a)~\empirical{39} packages
that vendor vulnerable libraries in their latest version,
and (b)~\empirical{312} client packages
whose code can reach vulnerable native functions
introduced
by their Python dependency chains.
The affected cases
account for over 47 million monthly downloads.
We responsibly disclosed all
findings to package maintainers,
and
{\bf 54} issues have been
fixed to date.

\point{Contributions}
Our work makes the following contributions:
\begin{itemize}[leftmargin=*,noitemsep,topsep=3pt]
\item We introduce a novel approach that resolves 
	the provenance of vendored libraries to specific
	OS package versions or built-from-source upstream versions.
    Our approach is implemented as a tool
    that we call~\tool.
\item  We evaluate~\tool\ by assessing its ability to resolve
    the provenance of vendored libraries
    and by comparing it against existing approaches.
    \tool\ correctly resolves provenance for 216 of 217 binaries
    and achieves 100\% CVE precision with 99.5\% recall, while the
    four state-of-the-art baselines achieve at most 32.9\% precision.
\item We integrate our approach with existing
    call graph generators and perform the first,
    cross-ecosystem
    vulnerability analysis
    for popular Python packages.
    We find hundreds of vulnerable packages,
    analyze their root causes,
    and propose ecosystem-wide
    hardening practices.
\end{itemize}

\section{Background and Terminology}
\label{sec:background}
\begin{figure*}[t]
\centering
\begin{subfigure}[t]{0.46\textwidth}
\vspace{0pt}
\begin{lstlisting}[language=Python,basicstyle=\scriptsize\ttfamily,numbers=left,numberstyle=\tiny,stepnumber=1,breaklines=true,xleftmargin=1.5em]
# .github/workflows/build.yml
- name: Build wheels (manylinux)
  uses: pypa/cibuildwheel@v3.3.0 # invokes auditwheel
  env:
    CIBW_BEFORE_BUILD: "yum install -y libxml2-devel && python setup.py"
    CIBW_BUILD: "*-manylinux_x86_64"
# igraph C comp.: etc/cmake/pkgconfig_helpers.cmake
set(PKGCONFIG_REQUIRES "libxml-2.0") # adds -lxml2 dep
\end{lstlisting}
\caption{The CI workflow installs {\tt libxml2-devel}
inside a {\tt manylinux} container, and {\tt igraph} declares
{\tt libxml-2.0} as a CMake dependency, making
{\tt \_igraph.abi3.so} link against {\tt libxml2}.}
\label{fig:igraph-yaml}
\end{subfigure}
\hfill
\begin{subfigure}[t]{0.52\textwidth}
\vspace{0pt}
\centering
\includegraphics[width=\linewidth]{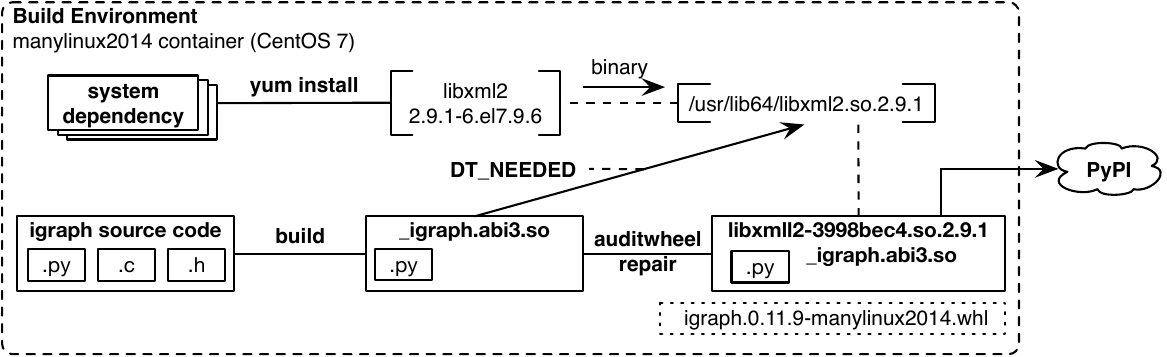}
\caption{Once the native extension is built,
PyPA's {\tt auditwheel}~\cite{auditwheel} resolves its shared-library
dependencies in the build environment,
and copies each into the wheel.
}
\label{fig:auditwheel-workflow}
\end{subfigure}
\vspace*{-3pt}
\caption{How {\tt libxml2} from a RedHat package ends up vendored inside the {\tt igraph}
	Python wheel. (a) The build configuration of {\tt igraph} links its native extension against
{\tt libxml2}, and (b) {\tt auditwheel} then copies {\tt libxml2.so} into
the final wheel.}
\label{fig:igraph-build}
\end{figure*}


\point{Python package distribution}
Python packages are published
as source distributions (sdists)
or~\textit{wheels},
which are pre-built artifacts~\cite{pep427}
that may bundle Python code
with compiled native components.
Among the latter we distinguish
\textit{native extensions} (native modules co-developed with the
Python code) from other binaries, which include both project-internal
auxiliary binaries and third-party libraries from external upstream
projects. We treat a binary as a third-party vendored library if it is
available in an OS distribution repository.
Since {\tt pip},
the Python package manager,
installs wheels by default when available~\cite{pypa-installing},
most deployments rely on these pre-built artifacts~\cite{python-survey-2024}.
The \textit{manylinux} specification~\cite{pep513}
restricts which system libraries wheels may link against to
ensure portability, and the
\texttt{auditwheel} tool~\cite{auditwheel} enforces this by
\emph{vendoring},
i.e.,
copying external shared libraries
into the wheel. Using \texttt{auditwheel} is not mandatory:
roughly one in two Python packages that ship third-party
binaries adopt it, while the rest copy binaries into the wheel
through custom build scripts.

Figure~\ref{fig:igraph-build} illustrates this for \texttt{igraph}.
Its CI workflow~\cite{cibuildwheel} installs
\texttt{libxml2} inside a \texttt{manylinux} container and links
the native extension against the installed \texttt{libxml2}
(Figure~\ref{fig:igraph-yaml}).
It then invokes \texttt{auditwheel}, which copies each shared-library
dependency into the wheel and patches the \texttt{DT\_NEEDED} and
\texttt{RPATH} entries of the \texttt{.so} files to reference the
bundled copies~\cite{auditwheel-repair} (Figure~\ref{fig:auditwheel-workflow}).
%
This achieves an effect similar to static linking,
trading central maintenance for portability.
PEP~513 acknowledges the tradeoff:
wheels bundling libraries like OpenSSL
\textit{``assume responsibility for prompt updates''}~\cite{pep513}.

\point{OS package distribution}
Operating-system (OS) distributions such as Debian GNU/Linux and RHEL
ship software in units called packages.
We refer to these as
\emph{OS packages} to distinguish them from Python packages.
The sources of these packages
are commonly referred to
as their \emph{upstream}
(e.g., OpenSSL's upstream is its GitHub
repository~\cite{openssl-github}, which Debian maintainers
package downstream~\cite{debian-openssl-salsa}).
A \emph{release} snapshots packages at specific upstream versions,
and over its lifetime
only security and
critical bug fixes are applied~\cite{debian-releases-wiki}.
Distribution maintainers are responsible for triaging
vulnerabilities and,
when necessary,
backporting fixes
from the upstream repository
to each version included in a supported release~\cite{debian-security-tracker}.
Therefore,
a library shipped in an OS distribution has two relevant versions:
(1) its~\textit{upstream version},
referring to the original project release,
and (2) its~\textit{OS package version},
typically an increasing integer number
that reflects distribution-specific updates over time.
For example,
Debian 9 includes OpenSSL (upstream)
version 1.1.0l as its \texttt{openssl} package,
and the latest version of the package for Debian 9
is \texttt{1.1.0l-1\textasciitilde deb9u11},
the eleventh revision, containing fixes for 25 CVEs since the upstream release.
%
%

%

\section{Provenance Analysis}
\label{sec:provenance}

\point{Challenges}
Given a specific vendored library
found in a Python package distribution,
determining its provenance is a challenging task
that involves two key challenges.

\textit{{\bf [C1]} Build procedures complicate provenance tracking.}
Many vendored libraries
are copied from popular
OS distributions (Debian, Red Hat),
but build procedures and Python's packaging pipelines modify the binary,
breaking direct byte-level matching
against the original
(Section~\ref{sec:background}).
In less common cases,
developers manually build vendored libraries
from upstream source code instead of
reusing OS packages.

\textit{{\bf [C2]} Upstream version identification is an open problem.}
When maintainers
build libraries from source,
identifying the exact upstream version of library binaries
is necessary.
Existing approaches are inadequate for this task.
Ad-hoc approaches based on extracting
strings from library binaries are error-prone and
cover only a fraction of cases
(Section~\ref{sec:upstream-project-resolution}),
while similarity-based approaches~\cite{OSSPolice,LibvDiff,DeeperBin,insight}
are not designed for differentiating
between versions which differ only in bug fixes
(see our comparison in Section~\ref{sec:eval-comparison}).

\begin{figure}[t]
\centering
\includegraphics[width=\columnwidth]{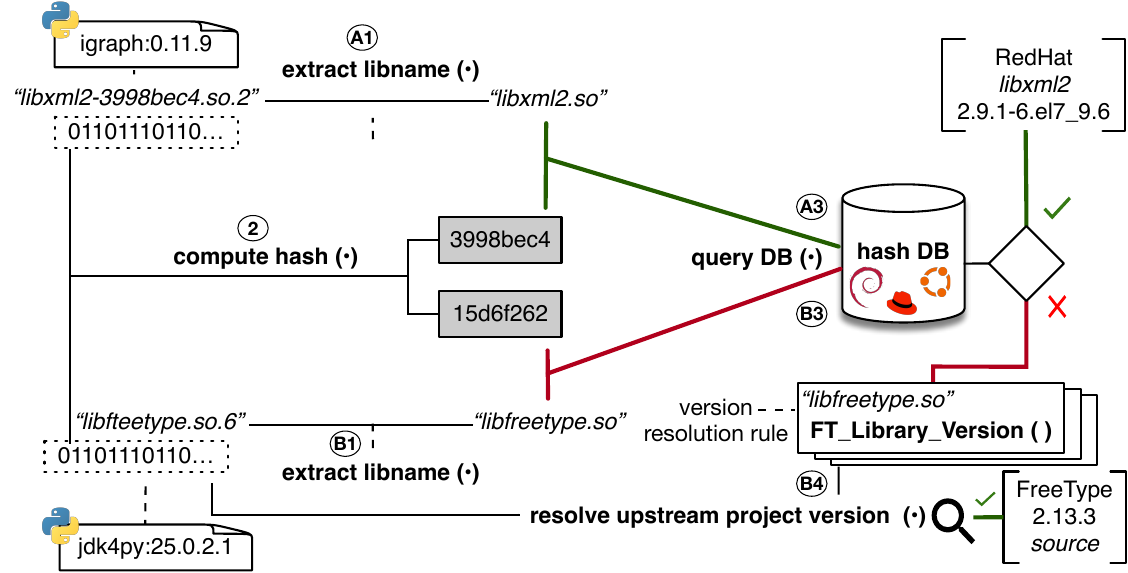}
\caption{Provenance resolution on two vendored libraries.
\calloutfigure{A} {\tt libxml2.so} from {\tt igraph}:
the binary hash matches our OS-package database,
resolving to
Red Hat {\tt libxml2}\textasciitilde2.9.1-6.el7\_9.6.
\calloutfigure{B} {\tt libfreetype.so.6} from {\tt jdk4py}: the binary's
SHA-256 finds no match, so we apply our library-specific rule to
recover {\tt FreeType}\textasciitilde2.13.3 built from source.}
\label{fig:provenance-analysis}
\end{figure}

To address the challenges above,
we employ two complementary procedures,
shown in Figure~\ref{fig:provenance-analysis}.
When a vendored library is copied from an OS package,
procedure \callout{A} resolves it to the specific OS package and version
(Section~\ref{sec:os-package-resolution}),
and when it is built from source,
procedure \callout{B} resolves it to the upstream project and version
(Section~\ref{sec:upstream-project-resolution}).
Each of these procedures relies on a key observation,
explained below.

\subsection{Resolution to OS Package and Version}
\label{sec:os-package-resolution}

Prior work~\cite{insight} resolves
the provenance of vendored libraries
by identifying their upstream project versions.
Vendored libraries in Python are often
(we quantify this in Section~\ref{sec:eval-provenance})
sourced from
OS distribution packages
that apply additional patches on top of
upstream versions to fix bugs and vulnerabilities.
Consequently,
two binaries built from the same
upstream version may differ from each other
when they originate from different OS package versions.

\point{Key observation}
Vendored libraries included in Python wheels
are often copied
from OS distribution packages.
Standard packaging pipelines in Python preserves
enough of the original binary to identify it.
For example,
{\tt auditwheel}
(Section~\ref{sec:background}) relocates
a vendored library and rewrites its ELF metadata
(e.g., the {\tt RPATH}
and {\tt DT\_NEEDED} entries).
However,
the executable
({\tt .text})
and read-only data
({\tt .rodata}) segments are left byte-for-byte
identical to the OS-package original. 
Based on this observation,
our method performs the following steps:

\point{Step 1: extract libname($\cdot$)}
Given the binary file of a vendored library,
we extract the~\emph{base} name of each vendored library
by removing any suffix from its file name.
This step is necessary
because vendored libraries are often renamed during packaging.
Recovering the base name allows us to match
the vendored binary
against candidate OS packages
that provide libraries with these base names
(see step 3).
For example,
the name of
{\tt libxml2-3998bec4.so.2.9.1} is normalized
to {\tt libxml2.so}
(\calloutt{A1}, Figure~\ref{fig:provenance-analysis}).

\point{Step 2: compute hash($\cdot$)}
We obtain a content-based identifier for
a vendored binary by hashing the segments
(\calloutt{2}, Figure~\ref{fig:provenance-analysis})
that packaging tools (e.g., {\tt auditwheel})
leaves unmodified.
This includes the {\tt .text} and {\tt .rodata}
of the binary as explained in our key observation.
We therefore compute a SHA-256 hash over these preserved segments.

\point{Step 3: query hash DB($\cdot$)}
Using our the hash
and the base name of each vendored library
computed from the previous steps,
we query a~\textit{hash database} of
OS package-provided shared libraries
(\calloutt{A3}, Figure~\ref{fig:provenance-analysis}).
This database contains entries for shared libraries
shipped in historical versions of major OS distributions
(e.g., Debian, RH, Ubuntu).
We defer the implementation details of how this database
is constructed to Appendix~\ref{sec:implementation-full}.
Each database entry corresponds to
a specific shared library instance
and records (1) the OS package name and version that provide it,
(2) the base name of the shared library,
and (3) the SHA-256 hash
of its binary contents.
For example,
an entry may indicate that
the library {\tt libxml2.so}
(hash prefix {\tt 3998bec4}) is provided by
the Red Hat package {\tt libxml2}
version 2.9.1-6.el7\_9.6.
By matching both the base name
and the hash of a vendored binary
against this database,
we find the specific OS package
version from
which the vendored binary originates.
To mitigate hash collisions,
we require the base library name
of the matched OS package to exactly match
the queried vendored library name before accepting a result.



\subsection{Resolution to Upstream Project Version}
\label{sec:upstream-project-resolution}

\point{The upstream version identification problem}
When a vendored library produces no hash match in our database
(e.g., {\tt libfreetype.so}
in Figure~\ref{fig:provenance-analysis}--\calloutt{B3}),
we conjecture that it was compiled directly from upstream source
rather than copied from an OS distribution package.
This conjecture does not widely hold in ecosystems such as Android or
embedded firmware,
where manufactures and downstream maintainers routinely
apply custom security patches on top of upstream
code~\cite{zhang2021androidkernel,fiber,capture,v1scan}.
The PyPI ecosystem is different: individual package maintainers
are unlikely to apply their own security patches for the
third-party libraries they vendor.
Under this assumption, the provenance of a binary without an
OS-package match is fully determined by its upstream project and
version.
We empirically validate this conjecture in
\S\ref{sec:eval-upstream-version}.

Extending the hash database to cover this case is impossible:
build non-reproducibility~\cite{benedetti2025reproducible,fourne2023flossing}
means that the same source produces binaries with different
hashes across compilations.
The goal of our fallback procedure
(when there is no hash match)
is thus to identify,
given a vendored
binary, its specific~\emph{upstream project release}.

\begin{table}[t]
\caption{Manually constructed version resolution
    rules for the 50 most frequently vendored libraries.
    \emph{Delimited} strings carry surrounding context in
    \texttt{.rodata} (e.g., "libcurl/8.19.0") that enables
    reliable regex matching: these are the only cases that prior
    \texttt{strings}/\texttt{grep}-based approaches handle.
    \emph{Bare} strings are plain \texttt{"X.Y.Z"} values
    indistinguishable from other version-like data.
    \emph{Opaque} values encode versions as packed integers
	or structs.
	\emph{Global variables} are accessible through dlopen/dlsym.
    \emph{Causal regex} targets delimited strings not                                                                                                
    exported as global symbols but defined as static variables                                                                                            
    in upstream source.
	}  
  \label{tab:rules}
  \scalebox{0.88}{%
  \begin{tabular}{lllrr}
    \toprule
    \textbf{Technique} & \textbf{Value} & \textbf{Greppable?} & \textbf{Libs} & \textbf{\%} \\
    \midrule
    Fn call $\rightarrow$ \texttt{char*}  & delimited  & \cmark & 8  & 16\% \\
    \rowcolor{gray!15}
    Fn call $\rightarrow$ \texttt{char*}  & bare       & \xmark & 12 & 24\% \\
    \rowcolor{gray!15}
    Fn call $\rightarrow$ int/struct      & opaque     & \xmark & 17 & 34\% \\
    \rowcolor{gray!15}
    Global var (\texttt{char*})      & bare       & \xmark & 3  & 6\%  \\
    \rowcolor{gray!15}
    Global var (packed int)         & opaque     & \xmark & 1  & 2\%  \\
    Causal regex                          & delimited  & \cmark & 4  & 8\%  \\
    \midrule
    \emph{No version info}                & ---        & ---    & 5  & 10\% \\
    \midrule
	  \multicolumn{3}{l}{\textbf{Total}}                          & \textbf{50} & \textbf{100\%} (\colorbox{gray!15}{66\%}) \\
    \bottomrule
  \end{tabular}}%
\end{table}

\point{Existing work and the gap}
Existing binary SCA work treats the version
identification problem
as a \emph{code-similarity}
problem, extracting generic features (string literals, CFGs,
function embeddings) and matching against a corpus of candidate
versions~\cite{JAHTNWZ24,b2sfinder,OSSPolice,LibvDiff,ATVHUNTER,DeeperBin}.
These approaches are heavyweight and imprecise
(Section~\ref{sec:eval-comparison}).
Several of these works~\cite{LibvDiff,insight,payer2021librarian}
observe that binaries sometimes contain explicit version information,
occasionally recovered via regex matching on strings included in the binary.
None, however, characterize how prevalent
such explicit version information is across real-world libraries.
We address this gap with the study below,
which guides the design of our solution.

\begin{figure}[t]
\begin{subfigure}{\columnwidth}
\begin{lstlisting}[language=c,basicstyle=\scriptsize\ttfamily,numberstyle=\scriptsize]
void *h = dlopen(argv[1], RTLD_NOW);
png_get_libpng_ver_t png_get_libpng_ver =
  dlsym(h, "png_get_libpng_ver");

const char *ver = png_get_libpng_ver(NULL);
printf("%s\n", ver);
\end{lstlisting}
\caption{{\tt libpng16.so}}
\label{fig:libpng}
\end{subfigure}

\vspace{0.5em}

\begin{subfigure}{\columnwidth}
\begin{lstlisting}[language=c,basicstyle=\scriptsize\ttfamily,numberstyle=\scriptsize]
void *h = dlopen(argv[1], RTLD_NOW);
FT_Library_Version_t FT_Library_Version =
  dlsym(h, "FT_Library_Version");
FT_Library lib;
FT_Init_FreeType(&lib);

int major, minor, patch;
FT_Library_Version(lib, &major, &minor, &patch);
printf("%d.%d.%d\n", major, minor, patch);
\end{lstlisting}
\caption{{\tt libfreetype.so}}                                                             
\label{fig:libfreetype}                                                                                                                                       
\end{subfigure}                                                                                                                                               
	\caption{Dynamic version resolution rules generated by our approach for {\tt libpng} and
	{\tt FreeType}.}
\label{fig:example-rules}
\end{figure}

\point{Exploratory study on version encodings}
To understand how libraries encode version information at the
binary level,
we manually inspect the source code and
sample binaries for the 50~most frequently
vendored libraries
and write a \emph{version resolution rule}
for each:
a small routine that takes the path to a
binary as input and outputs its upstream version.
For each library,
we inspect its source code and documentation to locate
where the version is defined
and how it is exposed at the binary level,
such as an exported function,
a global variable,
or a string literal embedded in the binary.
To confirm our reading,
we check out the repository tag for a known release
and trace the corresponding version value into a sample binary.
Each rule is validated against binaries with a known upstream version
yielding 50 rules in total.

Table~\ref{tab:rules} classifies the mechanisms we find into
five categories,
and three findings emerge.
First, two-thirds of libraries (66\%) encode their version in a
form~\emph{not} discoverable via static string matching (highlighted
rows).
This includes numeric strings
indistinguishable from other data,
integers accessible only
through function calls, or global variables requiring pointer
dereference. The regex-based extraction discussed by prior
work~\cite{payer2021librarian,LibvDiff} can recover only the 24\% of
libraries with version strings.
Second, 10\% of the top-50 libraries embed no version information
at all.
For example,
\texttt{libuuid.so} (from \texttt{util-linux})
declares its version only in \texttt{configure.ac},
which the build
system never compiles into the binary.
Third, even among libraries with callable version APIs,
correct extraction requires project-specific knowledge as integer
encodings vary across projects. For example, \texttt{libwebp}
packs three components into a single integer:
\smallskip\begin{lstlisting}[language=C,basicstyle=\scriptsize\ttfamily,numbers=none,xleftmargin=1em,aboveskip=2pt,belowskip=2pt]
int WebPGetEncoderVersion(void) {
  return (MAJ_VER << 16) | (MIN_VER << 8) | REV_VER;
}
\end{lstlisting}

\point{Key observation}
Most libraries embed their own version in the binary,
but rarely as a plain,
greppable string.
As our study shows,
most of the libraries expose it only through mechanisms
that static matching cannot handle (e.g., callable interfaces),
and correct extraction is project-specific.
This motivates resolving the upstream version using small,
per-library rules that query each binary through
its version-reporting interfaces.
Doing this by hand
for hundreds or thousands of libraries is impractical.

\point{LLM-assisted rule generation}
Following recent work that uses LLMs to parse project
documentation and source code to synthesize executable
artifacts~\cite{crossible,promefuzz}, we design a prompt that
encodes the taxonomy and priority ordering of our study (Table~\ref{tab:rules})
into instructions for an LLM
(Claude Opus~4.7; the full prompt is shown in
Figure~\ref{fig:enhanced-prompt} of Appendix~\ref{appendix:prompts}).
For each
library, we provide the LLM with the library name, the upstream
project, and one sample vendored binary.
Note that we never disclose the
binary's expected version.
The prompt directs the model to (1)~study the upstream project's
source code to identify the version API; (2)~inspect the sample
binary to verify symbol availability; and (3)~write a rule using
the highest-priority technique that works, preferring dynamic
approaches (function calls, variable reads)
over static byte-pattern matching.
Fragile techniques such as generic \texttt{strings|grep},
SONAME parsing, and \texttt{.comment}-section reading are
explicitly rejected.

We validate each generated rule against a set of binaries with
known upstream versions,
taken from two sources:
(1)~\emph{all} instances of the library in our dataset
whose provenance resolves to a specific OS package version via
exact hash matching~(Section~\ref{sec:os-package-resolution}),
and (2)~if fewer than five such instances exist,
we supplement with up to five
additional libraries extracted directly from
Debian, Red Hat, and Ubuntu packages in our hash database.
In both cases,
the ground-truth upstream version is recovered
from the OS package version string.
For example,
the Red Hat package {\tt libxml2} at version {\tt 2.9.1-6.el7\_9.6} maps to
upstream project version {\tt 2.9.1}.
The validation set is~\emph{disjoint}
from the binary shown to the LLM
during rule generation.
We accept a rule if it produces the expected version on
every validation sample.

\point{Examples}
Figure~\ref{fig:example-rules} shows the rules we generate for
{\tt libpng} and {\tt FreeType},
two libraries that prior work~\cite{LibvDiff}
wrongly claims lack recoverable version information in their binaries,
using them to motivate similarity-based approaches.
For {\tt libpng}
(Figure~\ref{fig:libpng}),
scanning the binary for version strings is unreliable,
as the binary contains {\tt zlib}'s version string
as well as {\tt libpng}'s own,
so a string-matching technique~\cite{payer2021librarian} cannot tell
which belongs to {\tt libpng}.
Our rule avoids this ambiguity by
calling {\tt png\_get\_libpng\_ver()},
which returns {\tt libpng}'s version.
For {\tt FreeType} (Fig.~\ref{fig:libfreetype}),
LibvDiff~\cite{LibvDiff} claims
no version information is present in the binary,
and indeed there is no version \emph{string},
because {\tt FreeType} stores its version as three
integers rather than a string. Our rule recovers them by
initializing the library and invoking
{\tt FT\_Library\_Version()}.

Our analysis leverages these rules
generated by LLMs to resolve the provenance
of vendored libraries with no hash match.
As a concrete instance,
consider {\tt libfreetype.so.6} vendored
by {\tt jdk4py} in the bottom half of
Figure~\ref{fig:provenance-analysis}.
No exact match is found in
the hash database (\calloutt{B3}),
and
applying our {\tt libfreetype.so} rule to the
binary resolves (\calloutt{B4})
the provenance to {\tt FreeType}~2.13.3 built
from source.

\point{Libraries without version information}
When a library embeds no version information, our pipeline
produces no rule and we report only the upstream project
identity (resolved via OS package metadata), without a version.
As our evaluation will show
(Section~\ref{sec:eval-upstream-version}), roughly 18\% of the
libraries fall into this
category. Extending coverage
to these libraries
by integrating similarity-based version
estimation~\cite{LibvDiff,DeeperBin,bindiff} as a
fallback is left for future work.

\section{Evaluation}
\label{sec:evaluation}

We ask
the following research questions:

\begin{enumerate}[label={\bf RQ\arabic*}, leftmargin=2.3\parindent]
\item How complete is \tool?
\item How accurate is \tool\ on non-matches?
\item How does \tool\ compare to existing work?
\end{enumerate}

\subsection{RQ1: Completeness of Provenance Analysis}
\label{sec:eval-provenance}

\begin{table}[t]
\centering
\setlength{\tabcolsep}{4pt}
\caption{
Identification coverage (\%)
for resolving the provenance of vendored libraries,
either to an OS distribution package (``Hash'')
or to an upstream project version,
when built from source (``Ver.'').
Results are reported for all instances
and for security-relevant instances
(``Sec.-rel.'').
Top X\% denotes the most
frequent library basenames accounting
for X\% of all observed instances.
}
\label{tab:rq1}
\begin{tabular}{lrrrrrr}
\toprule
& \multicolumn{2}{c}{Hash (pp)} & \multicolumn{2}{c}{Ver. (pp)} & \multicolumn{2}{c}{Total (pp)} \\
\cmidrule(lr){2-3}\cmidrule(lr){4-5}\cmidrule(lr){6-7}
Scope & All & Sec.-rel. & All & Sec.-rel. & All & Sec.-rel. \\
\midrule
All
& 47.2 & 55.3
& 26.2  & 39.6
& 73.4 & \textbf{94.9} \\
Top 25\%
& 83.3 & 88.6
& 11.8 & 9.1
& 95.1 & \textbf{97.7} \\
Top 50\%
& 71.3 & 77.3
& 19.9 & 19.0
& 91.2 & \textbf{96.3} \\
Top 75\%
& 56.9 & 66.0
& 26.2 & 30.1
& 82.4 & \textbf{96.1} \\
\bottomrule
\end{tabular}
\end{table}

We begin by evaluating the
completeness/coverage of \tool,
i.e.\ for what percentage of instances
of vendored native libraries \tool\
can resolve their provenance.

\point{Setup}
We evaluate \tool\
on the \empirical{\nnum{1878}} Python packages
which vendor native libraries out of the
top~\nnum{100000} most popular Python packages.
We separate instances where
\tool\ resolves provenance to an OS package
(\textit{Hash})
or to a specific version of an upstream project
(\textit{Ver.}).
We label a vendored library instance as
security-relevant
if its upstream project has at least one assigned CVE.
Under this criterion,
\nnum{903} of ~\nnum{2407} library base names
(corresponding to \nnum{9307} out of \nnum{14068} library instances)
are security-relevant.

\point{Results}
Table~\ref{tab:rq1} presents our results.
Library frequency is skewed:
the top 23 library base names account for 25\% of
instances (130 for 50\%, 550 for 75\%).
For security-relevant instances the concentration is stronger
(11 base names cover 25\%, 52 cover 50\%, 211 cover 75\%). This
skew motivates reporting coverage both overall and for the most
frequent basenames.
Our approach resolves provenance for \empirical{94.9\%} of
security-relevant instances and \empirical{73.4\%} across all
instances. Of the security-relevant total, \empirical{55.3\%}
resolve to a specific OS package version via exact hash matching,
and \empirical{39.6\%} resolve to an upstream project version via
our dynamic analysis rules. The rest (5.1\%) are not resolved
by \tool\ since no analysis rule exists.

\subsubsection {Hash matching}
Hash-based exact matching covers \empirical{55.3\%} of
security-relevant instances. This is not a limitation of
the mechanism but a property of the dataset: only about half of
vendored binaries are copied from OS packages;
the rest are built from upstream source during the Python packaging
process.
Coverage concentrates on commonly vendored libraries.
For the top \empirical{25\%} of basenames by frequency,
\empirical{88.6\%} of instances are hash-matched.
For the top
\empirical{50\%}, exact-match coverage is \empirical{77.3\%}.
Unresolved cases primarily involve low-frequency libraries,
typically built from source.

\subsubsection {Upstream Project Version}
\label{sec:eval-upstream-version}
Of the \nnum{903} security-relevant libraries
in our dataset, \nnum{805} have at
least one instance that misses our hash database and thus
requires version-based resolution.
Our LLM-assisted rule generation pipeline
produces a dynamic analysis rule
for \nnum{657} (\empirical{82\%})
at a total~\emph{one-time} cost of \empirical{\$77}.
The remaining
\empirical{148} (\empirical{18\%}) likely embed no version
information at all, a rate consistent with our earlier finding
that 10\% of the top-50 vendored libraries lack a
version-encoding mechanism.
These libraries account for
\nnum{5.1\%} of security-relevant instances.
We report an ablation of our LLM prompt against a naive baseline
in Appendix~\ref{appendix:ablation}.


\subsection{RQ2: Accuracy on non-matches}
When a vendored library is not resolved to an OS package,
\tool\ assumes that the shared object was built from the source code of
the upstream project
(leading conjecture in Section~\ref{sec:upstream-project-resolution})
and applies a library-specific resolution rule to recover the
upstream version.
We perform the following experiment to evaluate in combination
that (a) the origin is indeed upstream source,
and (b) the exact version recovered by \tool\ is correct.
%
\begin{table}[t]
\centering
\caption{
Comparison against state-of-the-art tools on a benchmark of 217 binaries.
``CVE Detection'' reports precision, recall, and F1 over the 777
ground-truth (binary, CVE) pairs (one binary may be affected by multiple CVEs).
LibvDiff (shaded row) is evaluated only on a subset of binaries,
and its recall is reported relative to that subset.
}
\label{tab:eval}
\centering
\scriptsize
\setlength{\tabcolsep}{3pt}
\begin{tabular}{lrrrr}
\toprule
 & & \multicolumn{3}{c}{\textbf{CVE Detection}} \\
\cmidrule(lr){3-5}
\textbf{Tool} & \textbf{Analyzed} & \textbf{Precision} & \textbf{Recall} & \textbf{F1} \\
\midrule
\insight~\cite{insight} & 217/217 & 32.9 & 87.3 & 47.8 \\
BinaryAI~\cite{JAHTNWZ24} & 189/217 & 30.8 & 64.7 & 41.7 \\
cve-bin-tool~\cite{cvebintool} & 92/217 & 30.9 & 48.6 & 37.8 \\
\rowcolor{black!8}
LibvDiff~\cite{LibvDiff} & 97/97$^*$ & 26.3 & 88.2 & 40.5 \\
\tool\ (this work) & 216/217 & 100.0 & 99.5 & 99.7 \\
\bottomrule
\end{tabular}%
\vspace{-3mm}
\end{table}

\point{Setup}
For the 10 most frequent security-relevant library basenames,
we sample up to 10 instances per basename where no exact hash is matched,
and establish provenance ground truth
by inspecting the associated Python package's
packaging procedure (e.g., wheel build scripts, CI configuration)
in its public source repository.
Instances with no public packaging procedure
are discarded and resampled. This yields 97
instances for manual validation, as {\tt libbz2.so} has only seven instances.
%
The labeling was performed by an experienced security researcher
and checked by a second author.
For each instance, we record the exact lines
in the package's public packaging procedure
that establishes its provenance and version,
so every label can be verified directly against the source.

\point{Results}
In all \empirical{97} cases, we find that
the binary is ultimately built
from upstream source, confirming the conjecture.
In all cases, we also find that the version
recovered by \tool\ is correct.
In \empirical{94}
cases,
the packaging workflow compiles the third-party library directly.
In the
remaining three,
it uses a package manager
(Conan~\cite{conan-package-manager}, miniconda)
that itself builds from source.


\subsection{RQ3: Comparison with Existing Work}
\label{sec:eval-comparison}

We compare \tool\ against four
related tools:
the most closely related tool \insight~\cite{insight},
two state of the art academic tools
(BinaryAI~\cite{JAHTNWZ24},
and LibvDiff~\cite{LibvDiff}),
as well as the widely used industry tool
\texttt{cve-bin-tool}~\cite{cvebintool}.
The three former tools are similarity-based,
i.e.\ they extract and compare features from binaries
against known reference versions, while
\texttt{cve-bin-tool} relies on string-based
version identification (via regex).
All these tools are provenance-agnostic.
We evaluate each tool
on its ability to identify
the known vulnerabilities (CVEs) that
affect a given shared library.

\point{Setup}
\insight\ is the only related tool
that targets native libraries in the Python
ecosystem.
Its dataset and implementation are not publicly available
and we got no response when contacting the authors.
Therefore, we construct a novel evaluation benchmark
from the 72 publicly available GitHub issues
reported in the \insight{} paper.
From each issue,
we extract
the Python package version
and vendored binary,
and download
the corresponding PyPI wheel
to recover the referenced binary.
%
We obtain 217 binaries
from 77 distinct library base names.
To establish each binary's ground-truth provenance,
and thus the set of CVEs it is actually affected by,
we employ \tool{}'s hash identification
and find 188 come from OS distribution packages,
recording their exact versions.
For the remaining 29,
we manually analyze the corresponding packages'
build pipelines following RQ2's methodology
to determine ground truth.
We then query Trivy's~\cite{trivy} CVE database with
each binary's provenance to obtain its affecting CVE set,
defining our ground truth.
This yields a set of 777
$\langle$binary, CVE$\rangle$ tuples,
where each binary may be affected
by multiple vulnerabilities.

We evaluate
each tool
using its standard interface.
We query BinaryAI
through its API,
and we execute
{\tt cve-bin-tool}
directly on each binary.
LibVDiff
requires a reference database
of source-built binaries.
Due to its high computational cost,
we select five upstream projects
and populate LibVDiff's feature database
with 311 distinct versions
of source-built binaries
from OpenSSL, krb5, libjpeg, libxml2 and util-linux.
We then evaluate LibVDiff
on the subset of 97 binaries
from our benchmark that match these projects.
Because the authors did not release
the original \insight{} implementation,
we extract
\insight{}'s reported upstream project version
for each binary
from the corresponding GitHub issues.
The claimed provenance for each of these tools
is that the binary was built from source
at the identified version.

Given the reported provenance of the binary from each tool,
we query Trivy's CVE database and obtain a set of
$\langle$binary, CVE$\rangle$ tuples.
We compare these tuples against the ground-truth
$\langle$binary, CVE$\rangle$ tuples.
Let $\mathcal{C}_{\mathrm{claimed}}$ and $\mathcal{C}_{\mathrm{actual}}$
denote the tool-reported and ground-truth sets of
$\langle b,c\rangle$ tuples, respectively.
Precision is the fraction of claimed CVEs actually applicable to the
binary, $\mathrm{Precision} = |\mathcal{C}_{\mathrm{claimed}} \cap
\mathcal{C}_{\mathrm{actual}}| / |\mathcal{C}_{\mathrm{claimed}}|$,
and recall is the fraction of actual vulnerabilities identified by the
tool, $\mathrm{Recall} = |\mathcal{C}_{\mathrm{claimed}} \cap
\mathcal{C}_{\mathrm{actual}}| / |\mathcal{C}_{\mathrm{actual}}|$.

Low precision indicates
that a tool frequently reports CVEs
that do not apply to the binary,
for example when it maps
a patched distribution package
to a vulnerable upstream release.

\begin{figure*}[t]
\centering
\includegraphics[width=\textwidth]{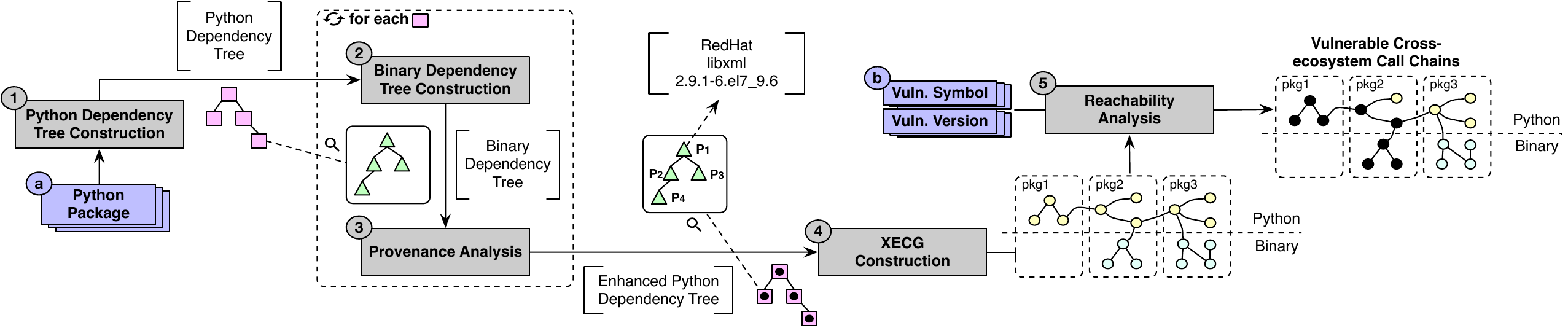}
\caption{
Our pipeline for
building cross-ecosystem call graphs (XECGs)
using \tool\
and performing reachability analyis.
}
\label{fig:arch}
\end{figure*}

\point{Results}
Table~\ref{tab:eval}
summarizes our results.
\tool{}
successfully analyzes
216/217 binaries.
The single failure is an instance of
{\tt libxerces.so} built from upstream source
at version 3.1.3 and affected by 4 CVEs.
The {\tt Xerxes-C} project,
from which {\tt libxerces.so} is built,
embeds no version information in the binary,
so \tool\ cannot create a version resolution rule for it.
Similarity-based techniques achieve high coverage,
with \insight{} (by the dataset's definition) producing outputs for 100\% of instances
and BinaryAI for 189/217 (87\%) of instances.
%
In contrast, {\tt cve-bin-tool},
which relies on string-based version identification,
produces output for only 42\% of instances.
This occurs because it requires a directly greppable version string
to identify a binary.


\tool{}
achieves
near-perfect recall,
missing only the {\tt libxerces.so} CVEs,
while producing no false positives.
In contrast,
all provenance-agnostic tools
exhibit low precision
(26.3--32.9\%)
despite relatively high recall
(48.6--88.2\%).
These tools infer vulnerabilities
solely from upstream releases,
and therefore frequently report CVEs
that do not apply to the binary.
In particular,
binaries originating from OS distribution packages
are often flagged as vulnerable
even when the distribution has
backported and applied the corresponding fixes.
Similarity-based approaches also
produce incorrect upstream version estimates in some cases,
which affects recall
when a patched version is inferred instead of the true vulnerable one.
Even for source-built binaries in our benchmark (29 cases),
these approaches struggle to recover the exact upstream version
(BinaryAI: 17/29, \insight{}: 6/29, LibVDiff: 5/14*).
Appendix~\ref{appendix:version-identification} reports results
exclusively evaluating upstream version identification,
and Appendix~\ref{appendix:old-insight-comparison} gives a
per-library breakdown of~\insight{}'s reports for the five most
popular affected packages.

Overall,
our results demonstrate that accurate binary vulnerability assessment
requires exact provenance recovery.
The limitations of existing tools stem from two issues:
(1) similarity-based version identification provides only an approximation,
and (2) ignoring OS distribution packages causes false positives
by missing backported fixes.


\section{End-to-end reachability analysis}
\label{sec:study}

In order to assess the prevalence
and impact of vulnerabilities
originating from vendored libraries
in the Python ecosystem, we integrate \tool\
with existing tools for call-graph construction
in the novel pipeline shown in Figure~\ref{fig:arch}.
%
The inputs to our pipeline are:
\callout{a} the Python package or application under analysis
and \callout{b} a set of
vulnerable binary symbols (function names),
derived from known vulnerabilities
and associated with specific OS package versions
or upstream project releases.
Deriving the set of vulnerable binary symbols
is orthogonal to our design and can be achieved
by existing SCA tools (e.g., Trivy~\cite{trivy})
and function identification approaches~\cite{wu2024function}.
Our pipeline outputs a set of~\emph{cross-ecosystem} call chains
that originate in Python,
traverse native and OS-level components,
and eventually reach vulnerable functions.
In the following,
we provide an overview of the main analysis steps.
More details are available in Appendix~\ref{sec:implementation-full}.

\begin{table*}[t]
\centering
\caption{Directly vulnerable Python packages.
Each row contains the:
(1) CVE,
(2) affected upstream project and library,
(3) vulnerable function and vulnerable upstream versions,
(4) \# of Python packages vendoring the library,
(5) \#  of packages for which we resolve provenance via a hash match,
(6) \#  of packages for which we resolve provenance
as upstream and extract their version via binary analysis,
(7) \#  of vulnerable Python packages with the upstream version criterion,
(8) \#  of vulnerable Python packages with our provenance criterion,
(9) false positive reduction incurred by our provenance analysis,
(10) \#  of Python packages that reach the vulnerable symbols
from their Python code,
and (11) \# of issues already fixed/reported.}
\label{tab:eval:rq3}
\begin{adjustbox}{max width=\textwidth}
\begin{tabular}{@{}l l l r r r r r r r r@{}}
\toprule
\textbf{CVE} &
\makecell[l]{\textbf{Upstream Project}\,/\\ \textbf{Library}} &
\makecell[l]{\textbf{Vuln.\ Function}\,/\\ \textbf{Vuln.\ Versions}} &
\makecell[c]{\textbf{\#}\\ \textbf{Vnd.}} &
\makecell[c]{\textbf{Hash}\\ \textbf{Match}} &
\makecell[c]{\textbf{Version}\\ \textbf{Match}} &
\makecell[c]{\textbf{\#\,Vuln.}\\ \textbf{Upstr.}} &
\makecell[c]{\textbf{\#\,Vuln.}\\ \textbf{Prov.}} &
\textbf{FP Red.} &
\makecell[c]{\textbf{\#}\\ \textbf{Reach}} &
\textbf{Fixed} \\
\midrule

\rowcolor{shaderow}
  & OpenSSL
  & \texttt{X509\_VERIFY\_PARAM\_set1\_policies}
  &
  & 166 & 75
  &
  &
  &
  &
  & \\
\rowcolor{shaderow}
\multirow{-2}{*}{CVE-2023-0464}
  & \texttt{libcrypto.so}
  & $<$1.1.1u;\,$<$3.0.9;\,$<$3.1.1;\,$<$1.0.2zh
  & \multirow{-2}{*}{241}
  & (69\%) & (31\%)
  & \multirow{-2}{*}{179/241}
  & \multirow{-2}{*}{102/241}
  & \multirow{-2}{*}{$-$43\%}
  & \multirow{-2}{*}{6}
  & \multirow{-2}{*}{3/6} \\
\addlinespace[2pt]

\multirow{2}{*}{CVE-2025-6021}
  & libxml2
  & \texttt{xmlBuildQName}
  & \multirow{2}{*}{50}
  & 34 & 16
  & \multirow{2}{*}{45/50}
  & \multirow{2}{*}{28/50}
  & \multirow{2}{*}{$-$38\%}
  & \multirow{2}{*}{6}
  & \multirow{2}{*}{5/6} \\
  & \texttt{libxml2.so}
  & $<$2.14.4
  &
  & (68\%) & (32\%)
  & & & & & \\
\addlinespace[2pt]

\rowcolor{shaderow}
  & FFmpeg
  & \texttt{avcodec\_receive\_frame}
  &
  & 2 & 51
  &
  &
  &
  &
  & \\
\rowcolor{shaderow}
\multirow{-2}{*}{CVE-2025-9951}
  & \texttt{libavcodec.so}
  & $<$8.0.0;\,$<$7.1.2;\,$<$5.1.7
  & \multirow{-2}{*}{53}
  & (4\%) & (96\%)
  & \multirow{-2}{*}{18/53}
  & \multirow{-2}{*}{17/53}
  & \multirow{-2}{*}{$-$6\%}
  & \multirow{-2}{*}{3}
  & \multirow{-2}{*}{1/3} \\
\addlinespace[2pt]

\multirow{2}{*}{CVE-2023-46218}
  & curl
  & \texttt{Curl\_cookie\_add}
  & \multirow{2}{*}{67}
  & 42 & 25
  & \multirow{2}{*}{37/67}
  & \multirow{2}{*}{8/67}
  & \multirow{2}{*}{$-$79\%}
  & \multirow{2}{*}{0}
  & \multirow{2}{*}{--} \\
  & \texttt{libcurl.so}
  & $\geq$7.46.0\,\&\&\,$<$8.5.0
  &
  & (63\%) & (37\%)
  & & & & & \\
\addlinespace[2pt]

\rowcolor{shaderow}
  & FreeType
  & \texttt{FT\_Load\_Glyph}
  &
  & 47 & 17
  &
  &
  &
  &
  & \\
\rowcolor{shaderow}
\multirow{-2}{*}{CVE-2025-27363}
  & \texttt{libfreetype.so}
  & $<$2.13.3
  & \multirow{-2}{*}{64}
  & (73\%) & (27\%)
  & \multirow{-2}{*}{54/64}
  & \multirow{-2}{*}{34/64}
  & \multirow{-2}{*}{$-$37\%}
  & \multirow{-2}{*}{3}
  & \multirow{-2}{*}{1/3} \\
\addlinespace[2pt]

\multirow{2}{*}{CVE-2024-33877}
  & HDF5
  & \texttt{H5T\_\_conv\_struct\_opt}
  & \multirow{2}{*}{61}
  & 28 & 33
  & \multirow{2}{*}{47/61}
  & \multirow{2}{*}{47/61}
  & \multirow{2}{*}{0\%}
  & \multirow{2}{*}{15}
  & \multirow{2}{*}{11/15} \\
  & \texttt{libhdf5.so}
  & $<$1.14.4
  &
  & (46\%) & (54\%)
  & & & & & \\
\addlinespace[2pt]

\rowcolor{shaderow}
  & PCRE2
  & \texttt{compile\_xclass\_matchingpath}
  &
  & 65 & 5
  &
  &
  &
  &
  & \\
\rowcolor{shaderow}
\multirow{-2}{*}{CVE-2022-1586}
  & \texttt{libpcre2-8.so}
  & $<$10.40.0
  & \multirow{-2}{*}{70}
  & (93\%) & (7\%)
  & \multirow{-2}{*}{61/70}
  & \multirow{-2}{*}{2/70}
  & \multirow{-2}{*}{$-$97\%}
  & \multirow{-2}{*}{1}
  & \multirow{-2}{*}{1/1} \\
\addlinespace[2pt]

\multirow{2}{*}{CVE-2020-10531}
  & ICU
  & \texttt{UnicodeString::doAppend}
  & \multirow{2}{*}{57}
  & 28 & 29
  & \multirow{2}{*}{34/57}
  & \multirow{2}{*}{10/57}
  & \multirow{2}{*}{$-$71\%}
  & \multirow{2}{*}{2}
  & \multirow{2}{*}{2/2} \\
  & \texttt{libicui18n.so}
  & $\leq$66.1
  &
  & (49\%) & (51\%)
  & & & & & \\
\addlinespace[2pt]

\rowcolor{shaderow}
  & LZ4
  & \texttt{LZ4\_decompress\_generic}
  &
  & 49 & 13
  &
  &
  &
  &
  & \\
\rowcolor{shaderow}
\multirow{-2}{*}{CVE-2021-3520}
  & \texttt{liblz4.so}
  & $\geq$1.8.3\,\&\&\,$<$1.9.4
  & \multirow{-2}{*}{62}
  & (79\%) & (21\%)
  & \multirow{-2}{*}{50/62}
  & \multirow{-2}{*}{18/62}
  & \multirow{-2}{*}{$-$64\%}
  & \multirow{-2}{*}{1}
  & \multirow{-2}{*}{1/1} \\
\addlinespace[2pt]

\multirow{2}{*}{CVE-2022-35737}
  & SQLite
  & \texttt{sqlite3\_*printf}
  & \multirow{2}{*}{31}
  & 6 & 25
  & \multirow{2}{*}{13/31}
  & \multirow{2}{*}{7/31}
  & \multirow{2}{*}{$-$47\%}
  & \multirow{2}{*}{2}
  & \multirow{2}{*}{1/2} \\
  & \texttt{libsqlite3.so}
  & $<$3.39.2
  &
  & (19\%) & (81\%)
  & & & & & \\

\bottomrule
\end{tabular}
\end{adjustbox}
\end{table*}

%
% ...

\callout{1}~Given a Python package/application in the form
of a {\tt .whl} file,
we construct a
\textit{Python dependency tree}
that captures all Python-level dependencies of the input package,
by recursively extracting dependency information from
Python wheels.
\callout{2}~Next, we expand each node in the Python dependency tree
into a ~\textit{binary dependency tree},
which captures binary files
included in the wheel 
along with their inter-dependencies.
To achieve this,
the tool
examines each binary's {\tt DT\_NEEDED} header
to determine the list of
dynamically linked libraries.
\callout{3}~Then, \tool{} performs~\textit{provenance analysis}
to identify the origin and version of each binary
(see Section~\ref{sec:provenance}).
%
The result is an~\textit{enhanced Python dependency tree}
that includes both Python- and binary-level dependencies,
with every binary explicitly tagged with
its provenance and version information.
\callout{4}~To construct the~\textit{cross-ecosystem call graph (XECG)}, this step employs state-of-the-art Python~\cite{pycg}
and binary~\cite{ghidra} call-graph generators
and stitches together the initially disjoint call graphs
using established stitching techniques~\cite{keshani,bloat-study,pyxray}.
The resulting XECG captures call edges across
the package and all of its Python- and binary-level dependencies,
as shown in the example of Figure~\ref{fig:run_ex}.
Note that call graphs for any given Python or binary program
are computed once, cached, and then stitched
on demand.
\callout{5}~Finally,
we perform a~\textit{reachability analysis} on
the constructed XECG to determine
whether the Python package
or application under test can reach a vulnerable symbol (function) in
any of its native dependencies.
A Python package or application is considered affected by a vulnerability
if the respective vulnerable symbol is reachable in the XECG.
%

\smallskip\par\noindent To demonstrate the practical
usefulness of \tool\ and assess
the impact of
vulnerable native libraries in the Python ecosystem,
we ask the following questions:
\begin{enumerate}[label={\bf RQ\arabic*}, leftmargin=2.3\parindent, start=4]
	\item Do popular Python packages
		vendor vulnerable native libraries?
\item Are popular Python packages
	transitively affected by vulnerabilities
	in native libraries?
\end{enumerate}

\subsection{RQ4: Directly Vulnerable Packages}
\label{sec:eval-direct}

\point{Goal}
We evaluate~the ability of our end-to-end pipeline
(Section~\ref{sec:study}) to identify
\emph{directly vulnerable} Python packages,
i.e.,
packages whose latest released wheels
vendor vulnerable native dependencies.

\point{Setup}
For each of the 10 most prevalent vendored libraries
observed in our dataset
(Section~\ref{sec:eval-provenance}),
we select one representative CVE satisfying two criteria:
(1) the vulnerability has been fixed in
at least one OS distribution package
(indicating it is significant enough to be backported),
and (2) the CVE metadata allows us to identify
the corresponding vulnerable symbol.
Using this information,
we analyze the top~\nnum{100000} most popular Python packages
and identify those whose
latest wheels bundle a vulnerable instance of the library.
We then construct the XECG for each candidate package
and perform reachability analysis to check
whether the vulnerable symbol is reachable from the package’s Python code.
Packages that satisfy both conditions are classified
as directly vulnerable.

\point{Results}
Our results are summarized in Table~\ref{tab:eval:rq3}.
Overall,
our pipeline detects~\empirical{39} Python packages
that vendor
library instances vulnerable to
at least one of the selected CVEs,
and the vulnerable function is
reachable from the package's Python code.
These include popular Python packages,
such as {\tt pymssql}
($>$35.3M monthly downloads)
and {\tt rasterio}
($>$3M monthly downloads).
Cumulatively,
the~\empirical{39} vulnerable Python packages have
$>$47M monthly downloads.
For each case, we further construct a~\emph{proof of
reachability}: a small Python program that calls the
package's public API and triggers a GDB breakpoint on
the vulnerable function.
We reported these issues to
package maintainers
following best practices,
and~\empirical{26}/\empirical{39} issues
have been fixed.

Our \tool-backed reachability analysis significantly reduces
false positives compared to a baseline approach
that ignores OS package versions
and backported fixes.
The average false positive reduction is~\empirical{48\%},
while it can reach up to~\empirical{97}\%
for libraries typically pulled
from OS packages
(e.g.\ \texttt{curl}, \texttt{PCRE2}).
For the HDF5 library,
we observe a~\empirical{0\%} reduction in false positives
because HDF5 instances are sourced
directly from the upstream project
and not from OS distribution packages.
Indeed,
both Debian and Red Hat have discontinued official HDF5 support
due to issues in the upstream development repository~\cite{debian-problems}.
The only other case where
provenance analysis does not significantly (>\,37\%)
reduce detections is \texttt{FFmpeg} (-\,6\%),
which according to
our results is typically built from source rather than
pulled from a distribution,
which is a direct result of the software not being
available in Red Hat
due to licensing and patent concerns~\cite{ffmpeg-patent}.


\subsection{RQ5: Transitively Vulnerable Packages}
\label{sec:eval-indirect}

\newcommand{\pkg}[2]{\texttt{#1}~\texttt{#2}}

\begin{table}[t]
  \centering
  \caption{Indirectly vulnerable Python packages.
  Each row contains the:
  (1) CVE,
  (2) a representative Python package $P$
  that vendors a vulnerable library,
  along with the affected package versions,
  (3) \#  of Python packages which depend on $P$ (dependents),
  (4) \#  of dependents which pin a vuln.\ version of $P$,
  (5) \#  of dependents that reach the CVE from their Python code,
  and (6) \#  of issues already fixed/reported.}
  \label{tab:eval:rq4}
  \renewcommand{\arraystretch}{1.4}
  \scalebox{0.80}{%
  \small
  \begin{tabular}{l l r r r c}
    \toprule
    \textbf{CVE ID} & \textbf{\makecell[l]{Package / \\ vuln.\ version}} & \textbf{\#rdeps} & \textbf{\#pin} & \textbf{\#reach} & \textbf{Fixed} \\
    \midrule
    \rowcolor{gray!15}
    2023-0464  & \pkg{pymssql}{$\le$ 2.3.10}    & 179   & 34   & 29   & 3/10 \\
    2025-6021  & \pkg{igraph}{$\le$ 0.11.9}     & 331   & 70   & 2    & 1/2  \\
    \rowcolor{gray!15}
    2025-9951  & \pkg{av}{$\le$ 15.1.0}         & 364   & 107  & 57   & 0/10 \\
    2023-46218 & \pkg{pycurl}{$\le$ 7.45.4}     & 160   & 56   & 44   & 10/10 \\
    \rowcolor{gray!15}
    2025-27363 & \pkg{pillow}{$\le$ 11.1.0}     & 16511 & 3424 & 104  & 5/10 \\
    2024-33877 & \pkg{netcdf4}{$\le$ 1.7.1}     & 1299  & 58   & 27   & 6/10 \\
    \rowcolor{gray!15}
    2020-10531 & \pkg{pyside6}{$\le$ 6.6.3.1}   & 848   & 41   & 32   & 2/10 \\
    2021-3520  & \pkg{tables}{$\le$ 3.8.0}      & 642   & 41   & 17   & 1/10 \\
    \rowcolor{gray!15}
    2022-35737 & \pkg{pymeos-cffi}{$\le$ 1.2.0} & 1     & 0    & --   & --   \\
    \bottomrule
  \end{tabular}}
\end{table}

\point{Goal}
We evaluate the ability of our pipeline
(Section~\ref{sec:study})
to detect transitively vulnerable Python packages.
A Python package {\tt A} is \emph{transitively vulnerable}
to a native library vulnerability when:
(1) {\tt A} depends on Python package {\tt B},
(2) {\tt A} pins (explicitly requires)
a version of {\tt B} as a dependency,
(3) the pinned version of {\tt B} is directly vulnerable
(see definition in Section~\ref{sec:eval-direct}),
and (4) the vulnerable symbol is reachable from
the Python code of {\tt A}.
For example, in Figure~\ref{fig:run_ex},
{\tt scppin} is transitively vulnerable via {\tt igraph}
to {\tt libxml2's}
CVE-2025-6021.

\point{Setup}
For each of the CVEs selected in RQ4
(Section~\ref{sec:eval-direct}),
we choose the most popular Python package
that vendors the affected library
and for which the vulnerable symbol is reachable.
We then identify all released versions of this Python package
that are affected by the CVE,
even if its latest version is not vulnerable.
Next,
we collect all client packages
that depend on this target package
and identify those that
(1) pin a vulnerable version
and (2) can reach the vulnerable symbol from their own Python code.

\point{Results}
Table~\ref{tab:eval:rq4} summarizes our results.
In total,
we find 312 transitively vulnerable Python packages.
Since analyzing each specific case
and submitting a detailed
report to maintainers is a resource-intensive task,
we have already submitted reports for the~\empirical{72}
most popular packages,
and are continuing the reporting process.
To date,
\empirical{28} issues have been fixed by developers
by upgrading the pinned version of their dependency.
The benefit of cross-ecosystem reachability analysis
is showcased by the significant reduction
in alerts when utilizing reachability information.
From~\empirical{\nnum{3831}} client packages
with a pinned vulnerable dependency,
only~\empirical{312} reach the vulnerable symbols.
This accounts for a 92\% reduction in noise.



\subsection{Root causes and ecosystem hardening}
\label{sec:eco-findings}

Our findings point to four recurring root causes, each with a
concrete mitigation. First, \emph{end-of-life build environments}:
when the OS release used for building reaches EOL it stops receiving
security patches (e.g., \texttt{manylinux2014} is based on CentOS~7,
EOL June 2024~\cite{centos7-eol}), so wheels built afterwards inherit
unpatched libraries and maintainers should migrate to a supported variant
such as \texttt{manylinux\_2\_28} (AlmaLinux~8, EOL
2029~\cite{almalinux8-eol}), or manually patch if they must stay on an
EOL image. Second, \emph{stale packages in pinned build
configurations}: pinning a {\tt cibuildwheel} version fixes the
container image tag~\cite{cibuildwheel-options} and \texttt{auditwheel}
copies whatever libraries it ships~\cite{auditwheel}, so without an
explicit \texttt{yum update} \texttt{before-all} step the wheel may
bundle known vulnerabilities.
Adding that update restores timely
patches at the cost of strict reproducibility~\cite{github-actions-security}.
Third, \emph{unattended libraries built from source}: compiling
vendored libraries directly shifts patch-tracking entirely onto
maintainers, who should monitor upstream CVE feeds or migrate to an
OS-sourced workflow. Fourth, \emph{incomplete SBOMs}: since
September~2025 \texttt{auditwheel} (used by 881 of our \nnum{1878}
packages) can emit SBOMs, but only for libraries copied from OS
packages~\cite{auditwheel-sbom}, omitting source-built binaries and
older artifacts and thus risking a false sense of completeness
This gap is also documented in the Java ecosystem~\cite{jbomaudit}.

\point{Key Takeaway}
\emph{More broadly, all projects shipping vendored
third-party code, regardless of how it is sourced, should establish a
policy for promptly rebuilding and releasing wheels when
vulnerabilities are disclosed.}
This is also acknowledged by Python package maintainers.
One of them stated after our disclosure:
\textit{``This is a good wake-up call
that wheels need to be recreated more regularly,
even when the actual (Python) project hasn't seen any updates,
because of possible vulnerabilities in vendored dependencies''}.

\subsection{Threats to Validity}


Our evaluation may not generalize beyond the libraries and packages we
study, but this risk is limited by construction: for RQ1 and RQ2 we
analyze every vendored native library across the top~\nnum{100000} PyPI
packages rather than a sample, and for RQ3 the 217-binary benchmark
(77 base names, mixing 188 OS-sourced and 29 source-built binaries) is
reconstructed from the independently reported GitHub issues of prior
work~\cite{insight}, so it is not biased in our favor. Reachability is
a proxy for risk and does not guarantee exploitability, which may
require attacker-controlled input or other preconditions and could be
assessed with testing-based approaches~\cite{chainfuzz}.
Moreover, our
static call-graph construction may overestimate reachable
vulnerabilities via infeasible paths or miss them due to dynamic
language features and indirect calls~\cite{static-recall}, though our
architecture is agnostic to the reachability method and can trade
precision for recall (e.g., via dynamic tracing).
Parts of our ground
truth (RQ2 and RQ3) rely on manual provenance classification, which might
contain errors: we mitigate this by having an experienced security
researcher label each case and a second author cross-check it,
recording the exact lines of each package's public build procedure so
that every label is independently verifiable against source.
Finally, although we focus on Python, the approach
applies to other ecosystems that use native libraries without
obfuscating them (e.g., Node.js, or R's Portable Linux Binary Packages,
whose distribution draws on Python's \texttt{manylinux/auditwheel}
workflow~\cite{posit2025rspm}), given careful attention to each
ecosystem's packaging conventions.

\section{Related work}
\point{Library version Identification}
Two main methods exist for identifying the
version of a given library from a binary:
either constructing rules to match
string literals in the binary,
or extracting features
(from source-code or binaries)
and calculating similarity metrics.
Generally, the first, simpler approach
is preferred when version information
is available in the binary.
There are several approaches
that use
string-matching~\cite{pandora, cvebintool},
feature-matching~\cite{JAHTNWZ24,DeeperBin,LibScout,LibID,LibvDiff,ATVHUNTER,LibScan},
or a combination of these methods
~\cite{OSSPolice,payer2021librarian},
mainly targeting library identification in
Android applications.
In contrast to the Android ecosystem,
where build tools like R8~\cite{R8}
and ProGuard~\cite{proguard}
may strip identifiers, shrink code, and flatten
package hierarchies, necessitating
obfuscation-resilient
detection research,
PyPI's vendoring pipeline preserves symbol tables,
sonames, and ELF structure largely intact.
This makes heavyweight semantic or graph-based matching
unnecessary in our setting.
In our system,
we utilize dynamic analysis
to handle cases such as
\texttt{freetype},
which stores its version
as three separate integers.
Our results
indicate that similarity-based approaches
are required for only \textasciitilde5.1\% of cases in the Python
ecosystem. 

\point{Patch Identification}
There exists a notable line of work
on the identification of patch presence
for binaries.
These approaches are generally
resource-intensive and complex
and are applied when
version information is unavailable or meaningless.
Much of this work is focused on
Android kernels, with
\emph{FIBER}~\cite{fiber} and \emph{PDiff}~\cite{PDiff},
as representative examples of approaches which
extract semantic signatures from fixes
and perform similarity-based matching.
Lares~\cite{Lares} performs semantic search,
also employing an LLM in several steps of the process,
and is shown to outperform
prior approaches~\cite{BinXray, Robin, PS3}.
We tested \lares~\cite{Lares}
on a dataset of 13 binary-CVE combinations
(5 most popular Python packages from our dataset).
We adapted it to use the latest
Claude Sonnet~4.6 as the LLM instead of
Sonnet~3.5 from the original paper.
%
\lares\ was applicable to 6 cases,
while in the other 7 the patched function was not exported from the vendored
shared object and \lares\ could not deliver a verdict.
On the applicable cases, \lares\ matched our ground truth in 4
and produced 2 false positives.
Overall, our observations suggest that
patch-presence testing is complementary to,
rather than a replacement for, provenance-aware analysis,
as it is heavyweight and assumes patch (fixing commit) knowledge and availability.

\point{Cross-language analysis}
Several works exist
on constructing cross-language
graphs between runtime-based programming
languages (Python, Javascript, Java)
and native components~\cite{pyxray,charon,gasket,jni2,polycruise}.
To the best of our knowledge,
these tools have only been used to
investigate native extensions,
i.e.\ native code that is part of the application,
and not third-party native code as we do.
In our implementation, we employ~\xray~\cite{pyxray}
for cross-language analysis.

\section{Conclusion}
Modern software increasingly relies
on cross-ecosystem dependencies.
The Python ecosystem exemplifies
this trend through its pervasive
use of vendored native libraries.
Our work demonstrates that
accurate vulnerability assessment
in such settings requires reasoning
about provenance,
i.e.,
not just
what version of a library is present,
but where it came from
and what patches it includes.
%
Our findings have immediate implications.
Package maintainers must recognize
that vendored dependencies
require active security attention.
Security tool developers must
account for backported
patches and vendored copies.
The broader lesson extends beyond Python:
as ecosystems increasingly interoperate,
security analysis must effectively bridge
their boundaries.

\bibliographystyle{IEEEtran}
\bibliography{main}

\appendices

\section{Implementation Details and Discussion}
\label{sec:implementation-full}

We have implemented our approach
as a command-line tool
in roughly 15K lines of Python code.
\tool\ takes as input a Python wheel
along with a set of vulnerable symbols
and vulnerable package/upstream project versions,
and implements all the steps described in
Sections~\ref{sec:provenance} and~\ref{sec:study}.
The input wheels are obtained either directly
from PyPI via the {\tt pip download} command,
when the target package is available there.
Otherwise,
they are locally generated for Python applications
not hosted on PyPI by building a wheel from
the application's setup configuration
(e.g., using {\tt pip install .}
or {\tt python setup.py}).
We next discuss key implementation details
and the limitations of~\tool.

\point{Call graph generators}
\tool\ employs PyCG~\cite{pycg}
for building the unified Python call graph,
and Ghidra~\cite{ghidra} for the binary one.
Other call graph generators could be used with minimal changes,
as this choice is an implementation detail.

\point{Constructing hash database}
\tool\ builds the hash database
(Section~\ref{sec:os-package-resolution})
using a lightweight,
lazy approach.
For each vendored library
identified in a Python package under analysis,
it extracts the base library name
and use OS package manager search utilities
(e.g., {\tt apt search} for Debian) to identify
OS packages that provide libraries with matching names.
These searches are executed inside Docker containers
running different OSes to account for variations in package naming,
library availability,
and package versions across distributions.

We currently restrict our analysis to packages
built for x86\_64 architectures,
which are supported by more than 99\% of Python wheels
that include shared libraries~\cite{manylinux-timeline}.
For each matching OS package,
\tool\ retrieves all historical versions,
downloads and extracts the binary package files
(e.g., {\tt .deb}, {\tt .rpm}) without installation,
computes the first 8 hexadecimal digits of the SHA-256 hash of
each contained shared library,
and adds an entry of the form
$\langle \mathit{os}, p, v, l, h \rangle$,
where $\mathit{os}$ indicates the operating system family,
$p$ the OS package name,
$v$ the package version,
$l$ the base name of a library included in $p$,
and $h$ the corresponding 8-hexadecimal-digit hash
of the library binary.
Table~\ref{tab:db-stats} summarizes
statistics about the contents of the hash database
used in our evaluation.

\point{Resolving the upstream project}
OS distribution packages from Debian and Red Hat include
machine-readable metadata that specifies the upstream project
from which each package is
derived~\cite{debian-copyright,rpm-spec}. Debian packages
({\tt .deb} files) carry a copyright specification with a
{\tt Source} field pointing to the upstream project
(e.g.,~a URL to its source repository); Red Hat packages
({\tt .rpm} files) carry the same information in their
{\tt spec} file. Given the base name of a vendored library
with no matching OS package version, our approach resolves its
upstream project by first identifying the binary OS package
that ships the library (e.g.,~via {\tt apt search} or
{\tt dnf provides}), then retrieving the corresponding source
package and reading its metadata to extract the upstream
project information. This procedure is invoked from the
LLM-assisted rule generation pipeline of
Section~\ref{sec:upstream-project-resolution}.

\point{Stripped binaries}
Reachability analysis
(Section~\ref{sec:study})
requires access to function-level information
and therefore cannot be performed on stripped binaries.
When encountering stripped
but potentially vulnerable shared libraries,
we substitute them with unstripped equivalents
before constructing the call graph.
For libraries originating from OS distribution packages
(Section~\ref{sec:os-package-resolution}),
we obtain unstripped versions
using the debugging symbol packages
provided by the distribution
(e.g., {\tt *-dbg} / {\tt *-dbgsym} packages in Debian-based systems).
For libraries that resolve to a specific upstream project version
(Section~\ref{sec:upstream-project-resolution}),
we rebuild the library from source with debug symbols enabled
before constructing the XECG.

\begin{table}[tp]
\centering
\caption{Summary statistics of packages
and package versions included in the hash database.}
\label{tab:db-stats}
\footnotesize
\setlength{\tabcolsep}{4pt}
\begin{tabular}{lllrr}
\toprule
\textbf{OS Family} & \textbf{Distribution} & \makecell{\textbf{Earliest OS} \\ \textbf{version}} & \textbf{\#pkgs} & \makecell{\textbf{\#pkg-} \\ \textbf{versions}} \\
\midrule
RedHat & AlmaLinux & 8.3 (2021)   & 1{,}136 & 5{,}041   \\
       & Fedora    & 12 (2009)    & 4{,}535 & 38{,}795  \\
       & EPEL      & 6 (2017)     & 1{,}964 & 6{,}140   \\
       & CentOS    & 3.1 (2004)   & 1{,}677 & 9{,}877   \\
       & Rocky     & 8.3 (2021)   & 1{,}090 & 4{,}184   \\
\midrule
\multirow{2}{*}{Debian} & Ubuntu & 14.04 (2014) & 5{,}459 & 20{,}035  \\
       & Debian    & 1.1 (1996)   & 7{,}907 & 241{,}519 \\
\bottomrule
\end{tabular}
\end{table}

\subsection{Limitations}
\label{sec:limitations-full}

\point{Exhaustiveness of hash database}
Our current implementation targets
ELF binaries on Linux.
While our approach could be extended to Windows libraries,
such binaries are less common on PyPI.
The effectiveness of our provenance analysis also depends on
the completeness of the hash database.
A fully exhaustive alternative would be to fetch
and analyze complete OS distribution mirrors
containing all historical package versions.
However,
this would require terabytes of storage and substantial processing.
Instead,
we construct the hash database on demand
(Appendix~\ref{sec:implementation-full}).
The database construction strategy is an implementation detail
and can be replaced by alternative mechanisms in the future.

\point{Custom repositories}
Another limitation of~\tool\ is
that it can currently resolve the provenance of vendored libraries
only when they have a corresponding OS package.
Even when a library is built from upstream source,
our approach relies on OS package metadata to identify
its upstream project.
Vendored libraries may instead originate from projects
that are not packaged in mainstream OS distribution repositories,
or are available only in custom repositories.
Some of these cases could be addressed by
extending the hash database to include
packages beyond the main repositories.

\section{LLM Prompts}
\label{appendix:prompts}

Figure~\ref{fig:enhanced-prompt} shows the full enhanced prompt
used in our LLM-assisted rule generation pipeline
(Section~\ref{sec:upstream-project-resolution}),
and Figure~\ref{fig:prompt-baseline} shows the baseline
prompt used in the ablation study of
Appendix~\ref{appendix:ablation}.
\begin{figure*}[tp]
\begin{tcolorbox}[colback=gray!5, colframe=gray!50, fontupper=\small\ttfamily, left=4pt, right=4pt, top=4pt, bottom=4pt]
\textbf{System:} You write small C programs that extract the upstream
project version from a vendored native shared library (ELF \texttt{.so}).
The program receives the path to a binary as \texttt{argv[1]},
loads it with \texttt{dlopen}, extracts the version, and prints it
to stdout (e.g., \texttt{1.2.11}). Exit 0 on success, non-zero on
failure. The rule must work on any vendored instance of the target
library, not only the example binary provided.

Your response MUST be a single JSON object and NOTHING else: no prose, no markdown fences, no explanation outside the JSON. Schema:

\{\\
\hspace*{1em}"rule\_body": "<full text of a C program>",\\
\hspace*{1em}"explanation": "<1-2 sentences explaining what the rule does>"\\
\}

\medskip
\begin{enumerate}[leftmargin=*, nosep]
\item \textbf{Study the upstream source code} of \{project\} to
  determine how the project exposes its version. Look for
  version-returning functions in public headers
  (e.g., \texttt{*\_version()}, \texttt{*\_get\_version()}),
  exported global variables
  (e.g., \texttt{*\_version}, \texttt{*\_VERSION\_STRING}), or
  static version strings embedded at build time. Note the exact
  function signature, return type, and integer encoding scheme
  where applicable.

\item \textbf{Inspect the binary} using \texttt{nm~-D},
  \texttt{readelf}, and \texttt{strings} to verify which version
  symbols are actually exported in the vendored binary.

\item \textbf{Write the rule} using the \textbf{first} technique
  that works, in priority order:

  \smallskip
  \begin{enumerate}[label=\alph*), nosep, leftmargin=*]
  \item \textbf{Call a version function that returns a string}
    (preferred): use \texttt{dlsym} to resolve an exported function
    that returns the version as \texttt{const char*}, then call it.

  \item \textbf{Call a version function that returns numeric data}:
    use \texttt{dlsym} to resolve a function that returns the
    version as a packed integer or fills integer output parameters.
    The exact encoding must be verified from the upstream
    source---do not guess.

  \item \textbf{Read a version symbol}: use \texttt{dlsym} to
    resolve an exported data symbol (\texttt{const char[]} or
    \texttt{const char*}) containing the version string.

  \item \textbf{Match a source-defined byte pattern} (only if
    a--c are impossible): read the raw binary and search for a
    version string that is defined as a \texttt{static} variable
    in the upstream source. Cite the source file and declaration.
    The string must have identifying context. Generic
    \texttt{X.Y.Z} patterns are not accepted.
  \end{enumerate}

  \smallskip
  Not accepted: generic \texttt{strings|grep}, SONAME/filename
  parsing, \texttt{.comment} section, symbol version tags.

\end{enumerate}

\medskip
\textbf{User:} \\
Library: \{libname\} \quad Upstream project: \{project\} \\
Path to example binary: \{path\_to\_binary\}
\end{tcolorbox}
\caption{LLM prompt template for version resolution rule generation.
  The technique priority order reflects our finding that dynamic
  approaches (function calls, global variable reads) account for
  82\% of the top-50 rules (Table~\ref{tab:rules}), while causal
  regex is reserved for the minority of libraries that embed
  delimited version strings without an exported API.}
\label{fig:enhanced-prompt}
\end{figure*}

\begin{figure}[tp]
\begin{tcolorbox}[colback=gray!5, colframe=gray!50, fontupper=\small\ttfamily, left=4pt, right=4pt, top=4pt, bottom=4pt]
\textbf{System:} You write small C programs that extract the upstream
project version from a vendored native shared library (an ELF
\texttt{.so} file). The rule must work on any vendored instance of
the target library, not only the example binary provided.

Your response MUST be a single JSON object and NOTHING else: no prose, no markdown fences, no explanation outside the JSON. Schema:

\{\\
\hspace*{1em}"rule\_body": "<full text of a C program>",\\
\hspace*{1em}"explanation": "<1-2 sentences explaining what the rule does>"\\
\}

The script receives the path to the binary as \texttt{argv[1]},
extracts the version, and prints it to stdout (e.g.,
\texttt{1.2.11}). Exit 0 on success, non-zero on failure.

\medskip
\textbf{User:} \\
Library: \{libname\} \quad Upstream project: \{project\} \\
Path to example binary: \texttt{\{path\_to\_binary\}}
\end{tcolorbox}
\caption{Baseline LLM prompt used in the ablation study
of Appendix~\ref{appendix:ablation}. Compared to the
enhanced prompt of Figure~\ref{fig:enhanced-prompt}, this version
omits the technique taxonomy, priority ordering, and explicit
guidance on what to study or reject.}
\label{fig:prompt-baseline}
\end{figure}

\section{Ablation of the LLM Prompt}
\label{appendix:ablation}

We ablate our LLM prompt against a baseline
(Figure~\ref{fig:prompt-baseline})
that lacks our empirically-guided instructions from the exploratory
study on version encodings
(Section~\ref{sec:upstream-project-resolution}).
Specifically, we consider 50~libraries ranked 51--100
by frequency, disjoint from the manually studied top-50
from Table~\ref{tab:rules};
ground truth on rule correctness
is established by manual rule construction using the
same methodology as in our exploratory study.
Of the 39~libraries that embed version information, the baseline
prompt produces valid rules in 19~cases, but 11 of those rely on
regex matches against bare version strings, which is fragile
across build environments.
Our enhanced prompt produces valid rules in 38/39 cases.
The single failure, \texttt{libhwloc.so}, requires a chain of
three function calls to recover
(Figure~\ref{fig:hwloc-rule}).

\begin{figure}[H]
\begin{lstlisting}[language=c, basicstyle=\small\ttfamily]
hwloc_topology_t topo;
hwloc_topology_init(&topo);
hwloc_topology_load(topo);

char *xmlbuf; int xmllen;
hwloc_topology_export_xmlbuffer(
    topo, &xmlbuf, &xmllen, 0);

/* parse hwlocVersion="2.2.0" from XML */
char *p = strstr(xmlbuf, "hwlocVersion=\"");
p += strlen("hwlocVersion=\"");
char *end = strchr(p, '"');
printf("%.*s\n", (int)(end - p), p);
\end{lstlisting}
\caption{Manually constructed version-extraction rule for
  \texttt{libhwloc.so}. The library offers no version-string
  function; \texttt{hwloc\_get\_api\_version()} returns only
  \texttt{major.minor} without the patch component. The full
  version is embedded as metadata in every topology object
  (\texttt{topology.c:3701}: \texttt{hwloc\_\_add\_info(\&topology->infos,
  "hwlocVersion", HWLOC\_VERSION)}) and is only accessible by
  initializing a topology, exporting it to XML, and parsing the
  \texttt{hwlocVersion} attribute.
  Both the baseline and enhanced LLM prompts failed to discover
  this three-step invocation chain, instead producing a
  fragile regex matching a bare \texttt{"2.x.y"} string in
  \texttt{.rodata}, a pattern that is not reliably
  distinguishable from other version-like data in the binary.}
\label{fig:hwloc-rule}
\end{figure}

\section{Upstream Version Identification}
\label{appendix:version-identification}

Table~\ref{tab:eval-version} reports each tool's ability to
recover the upstream version of a vendored binary, on the same
217-binary benchmark used in Section~\ref{sec:eval-comparison}.
For each tool we report how many predictions it produced,
how many exactly matched the ground-truth upstream version, and
the distribution of the signed version-index error
(distance in released versions between the predicted and the
ground-truth version, using the upstream project's release
history as the ordering).
LibvDiff (shaded row) is evaluated only on the subset of
binaries whose libraries are covered by its reference database.
\tool\ predicts the exact upstream version for every binary it
analyzes.

\begin{table}[tp]
\centering
\caption{Upstream version identification on the 217-binary
benchmark. ``Correct'' is the number of predictions that
exactly match the ground-truth upstream version. Median, Mean,
P5, and P95 report the signed version-index error (0 = exact
match; negative = older than ground truth).}
\label{tab:eval-version}
\scriptsize
\setlength{\tabcolsep}{4pt}
\resizebox{\columnwidth}{!}{%
\begin{tabular}{lrrrrrr}
\toprule
 & & & \multicolumn{4}{c}{\textbf{Version Skew}} \\
\cmidrule(lr){4-7}
\textbf{Tool} & \textbf{Predictions} & \textbf{Correct} & \textbf{Median} & \textbf{Mean} & \textbf{P5} & \textbf{P95} \\
\midrule
\insight~\cite{insight}         & 217 & 68  & 1 & -0.40 & -2.20 & 6    \\
BinaryAI~\cite{JAHTNWZ24}       & 189 & 81  & 0 & 1.52  & -2    & 8    \\
cve-bin-tool~\cite{cvebintool}  &  92 & 86  & 0 & -0.35 & 0     & 0    \\
\rowcolor{black!8}
LibvDiff~\cite{LibvDiff}        &  97 & 38  & 0 & -0.32 & -2    & 2    \\
\tool\ (this work)              & 216 & 216 & 0 & 0     & 0     & 0    \\
\bottomrule
\end{tabular}%
}
\end{table}

\section{Detailed Comparison with \insight\ on Issue Reports}
\label{appendix:old-insight-comparison}

Table~\ref{tab:insight} presents a per-library breakdown of the
comparison with~\insight\ discussed in
Section~\ref{sec:eval-comparison}.
The comparison covers the five most popular Python packages (by monthly
download count) for which the authors of~\insight~\cite{insight}
filed GitHub issue reports about vendored shared libraries.
Across these five packages, \insight\ reported a total of 13 vendored
vulnerable libraries.
For each instance, we list the upstream vulnerable version range
(per the NVD), the library version reported by~\insight, and the
provenance resolved by~\tool\ (i.e., the OS package and version,
with the upstream project version shown in bold).
We also indicate the vulnerability status of the library at the time
the issue was filed (``FIXED'', ``Vulnerable'',
or ``Unknown'').

\begin{table*}[tp]
\centering
\caption{
Analysis of~\insight\ reports
for the five most popular Python packages.
Each GitHub report may include
multiple vulnerabilities,
affecting 13 vendored libraries
in total.
For space reasons,
we list only the most recent CVE as the ``Indicative CVE''.
For each vendored library,
we report (1) the upstream vulnerable version range,
(2) the library instance version reported by~\insight,
and (3) the provenance resolved 
by our approach (OS package and version,
with the upstream project version shown in bold).
We also indicate the vulnerability status of each library
at the time of the report
(``FIXED'', ``Vulnerable'',
or ``Not Affected (N.A.)'').
Red fonts indicate cases
where~\insight\ reports
a wrong upstream project version.}
\label{tab:insight}
\resizebox{\textwidth}{!}{%
\begin{tabular}{@{} >{\ttfamily}l >{\ttfamily}l l l l >{\ttfamily}l l @{}}
\toprule
\normalfont\textbf{Project/Issue} & \normalfont\textbf{Library} & \normalfont\textbf{Ind. CVE} & \normalfont\textbf{Vuln. Upstr. v.} & \normalfont\textbf{Insight v.} & \normalfont\textbf{Actual Provenance (OS:package)} & \normalfont\textbf{Status before} \\ \midrule
pymssql/\#753 \cite{pymssql753} & libcrypto & 2021-3712 & $\le$1.1.1k & $\le$1.1.1 & D:libssl1.1/\textbf{1.1.0l}-1\textasciitilde deb9u4 & FIXED\\
triton/\#489 \cite{triton489} & libtinfo & 2021-39537 & $\le$6.2.1 & 5.9 & RH:ncurses-libs/\textbf{5.9}-14.20130511.el7\_4 & FIXED/N.A.\\
opencv/\#646 \cite{opencv646} & libcrypto & 2021-3712 & $\le$1.1.1k & \textcolor{red}{1.1.1f} & Upstream:\textbf{1.1.1g} & \textbf{VULN.}\\
psycopg/\#262 \cite{psycopg262} & libcrypto & 2020-7043 & $\le$1.1.1k & \textcolor{red}{1.1.0g} & D:libssl1.1/\textbf{1.1.0l}-1\textasciitilde deb9u5 & FIXED/N.A.\\
psycopg/\#262 \cite{psycopg262} & libidn & 2016-6262 & $\le$1.32 & \textcolor{red}{1.28} & D:libidn11/\textbf{1.33}-1+deb9u1 & FIXED\\
psycopg/\#262 \cite{psycopg262} & libp11-kit & 2020-29361 & $\le$0.23.21 & \textcolor{red}{0.23.2} & D:libp11/\textbf{0.23.3}-2+deb9u1 & FIXED\\
psycopg/\#262 \cite{psycopg262} & libgmp & 2021-43618 & $\le$6.2.1 (32-bit) & \textcolor{red}{$\le$6.1.0} & D:libgmp10/\textbf{6.1.2}+dfsg-1 & N.A.\\
psycopg/\#262 \cite{psycopg262} & libk5crypto & 2021-37750 & $<$1.19.3 & $\le$1.16 & D:libk5crypto3/\textbf{1.15}-1+deb9u3 & FIXED\\
psycopg/\#262 \cite{psycopg262} & libnettle & 2021-3580 & $<$3.7.3 & \textcolor{red}{$\le$3.2} & D:libnettle6/\textbf{3.3}-1+deb9u1 & FIXED\\
psycopg/\#262 \cite{psycopg262} & libsasl2 & 2020-8032 & $\le$2.1.27 (SUSE) & \textcolor{red}{2.1.26} & D:libsasl-2/\textbf{2.1.27}+deb9u1 & N.A.\\
psycopg/\#262 \cite{psycopg262} & libtasn1 & 2018-6003 & $<$4.13 & \textcolor{red}{4.7} & D:libtasn1-6/\textbf{4.10}-1.1+deb9u1 & FIXED\\
pygit2/\#1136 \cite{pygit21136} & libcrypto & 2021-23841 & $<$1.0.2y & 1.0.2k & RH:openssl-libs/\textbf{1.0.2k}-24.el7\_9 & FIXED\\
pygit2/\#1136 \cite{pygit21136} & libk5crypto & 2021-37750 & $<$1.19.3 & \textcolor{red}{1.16} & RH:krb5-libs/\textbf{1.15.1}-54.el7\_9 & FIXED\\
\bottomrule
\end{tabular}%
}
\end{table*}

\end{document}